\documentclass{article}

\usepackage{amsfonts}
\usepackage{amsmath}
\usepackage{amssymb}
\usepackage{graphicx}
\usepackage{geometry}
\usepackage{hyperref}
\usepackage{cite}
\providecommand{\U}[1]{\protect\rule{.1in}{.1in}}

\begin{document}

\title{Exact Solutions of the DKP Oscillator in 3D Spaces with Extended Uncertainty Principle}
\author{{Mokhtar Falek}$^{1}$ \and {Mustafa Moumni}$^{1,a}$ \and {Mahmoud Merad}$^{2}$ \\
$^{1}$Laboratory of Photonic Physics and Nano-Materials (LPPNNM)\\
Department of Matter Sciences, University of Biskra, ALGERIA \\
$^{2}$Faculty of Exact Sciences, University of Oum-el-Bouaghi, ALGERIA\\
$^{a}$correspondant author m.moumni@univ-biskra.dz}
\maketitle

\begin{abstract}
We present the exact solutions of the three-dimensional Duffin--Kemmer--Petiau oscillator for both spin 0 and spin 1 cases, with the presence of minimal uncertainty in momentum for anti--de Sitter model and we derive the solutions for the case of deSitter space. We use the representation of vector spherical harmonics and the method of Nikiforov--Uvarov to determine exactly the energy eigenvalues and the full expressions of the eigenfunctions in all cases. Our study of the energy spectrum allows us to define a new interpretation of natural and unnatural parity states of the vector particle and we show the crucial role played by the spin--orbit coupling in this differentiation between the parities.


P. A. C. S. 03.65.Ge, 03.65.Pm.
\end{abstract}

\section{Introduction}

During recent years, there has been a growing interest in absorbing the divergences appearing in quantum field theory (and other theories aimed at unifying the fundamental interactions) and several scenarios have been proposed to solve this kind of problems. Among these attempts, we find the extension of the quantum field theory to curved space-time, which can be considered as a first approximation of the theory of quantum gravity. In such situation, the usual Heisenberg uncertainty principle can be replaced by the so--called extended uncertainty principle (EUP) which is characterized mainly by the existence of a minimal length scale in the order of the Planck length \cite{Mignemi10}. A fundamental consequence deduced from this extension is that the minimal length uncertainty in quantum gravity can be related to a modification of the standard Heisenberg algebra by adding small corrections to the canonical commutation relations; we quote here Mignemi which has shown that, in a (anti--)de Sitter background, the Heisenberg uncertainty principle is modified by introducing corrections proportional to the cosmological constant \cite{Mignemi10}. Such a scenario is motivated by Doubly Special Relativity (DSR) \cite{Amelino01, Amelino02}, string theory \cite{Capozziello00}, non-commutative geometry \cite{Douglas01}, black hole physics \cite{Scardigli99, Scardigli03} and even from Newton's gravity effects on quantum systems \cite{Kuzmichev20}.

Recently, the introduction of this idea of EUP has drawn great interest and a significant number of papers appeared in the literature to address the effects of the extended commutation relations in quantum mechanics systems. We cite here the studies of thermodynamic properties of the relativistic harmonic oscillators on anti-de Sitter (AdS) space \cite{Hamil18}, the Klein--Gordon oscillator in an uniform magnetic field \cite{Chung19}, the exact solution of (1+1)-dimensional bosonic oscillator subject to the influence of an uniform electric field in AdS space \cite{Moussa18}. In addition, certain problems have also been solved in non-relativistic quantum mechanics despite the fact that, in conventional field theory approach of static de Sitter (dS)\ and AdS space--time models, we cannot derive any nonrelativistic covariant Schr\"{o}dinger--like equation from covariant Klein--Fock--Gordon equation. In this context, we can use the EUP formulation to write the dS and AdS versions of the Schr\"{o}dinger equation. Indeed, we find the treatment of the exact solution of the D--dimensional Schr\"{o}dinger equation for the free particle and the harmonic oscillator in AdS space \cite{Hamil19}, the study, with perturbative methods, of the implications of extended uncertainty principle of dS Space on the spectrums of both harmonic oscillator and hydrogen atom \cite{Ghosh11} and the exact solution of the Schr\"{o}dinger equation for the hydrogen atom in dS and AdS spaces \cite{Falek20}.

In this work, we solve the three--dimensional (3D) Duffin--Kemmer--Petiau (DKP) oscillator for spin 0 and spin 1 particles in AdS models as it was done very recently for the Dirac oscillator \cite{Merad19}. We will show rigorously that the problem admits analytical solutions for both scalar and vector particles; so we will compute the exact expressions of the eigenenergies and write the final forms of the eigenfunctions in all cases. Our interest in the DKP equation \cite{Duffin38, Kemmer39, Petiau36} comes from the fact that it is richer than those of Klein-Gordon and Proca and therefore it has more potential applications, especially for the study of hadrons and nuclei \cite{Clark85, Kalber86, Kozack89}. We also find studies on DKP in Hamiltonian covariant dynamic \cite{Kanatchikov00}, in Galilei covariance \cite{Montigny00} and in different topologies such as non--commutative spaces \cite{Falek08, Hassanabadi12, Achour19} or curved space--times \cite{Lunardi00, Castro15, Hosseinpoor18}. We focus on the harmonic oscillator given the great interest for this potential in quantum systems since it reflects a confinement with a non-zero residual energy. That is why it is the central potential of the nuclear shell model and also of the confining two--body potential for quarks. The relativistic version of the harmonic oscillator generated much interest especially since the work of Moshinsky and Szczepaniak \cite{Moshinsky89} for the Dirac equation and the works of Nedjadi and Barrett \cite{Nedjadi941, Nedjadi942} for the DKP version; we refer the reader here to the works already cited \cite{Hamil18, Chung19, Moussa18, Hamil19} or to \cite{Castro15} and \cite{Hosseinpoor19} where there is a very extensive list of references on both Dirac and DKP oscillators. We mention here that the relativistic oscillator was experimented recently \cite{Franco13, Fujiwara18}.

The outline of this paper is as follows: In the next section \ref{sec:RV}, we give a review on dS and AdS models, while in the third section \ref{sec:NU}, we introduce Nikiforov--Uvarov (NU) method used in our work to solve the system. In section four \ref{sec:S0}, we expose the explicit calculation of the deformed 3D DKP oscillator for spin 0 case in the framework of EUP; we do this in position space representation. By a straightforward calculation, our system will be converted to a Klein--Gordon equation type and we use the representation of vector spherical harmonics to solve it analytically; the corresponding radial wave functions are expressed with the Jacobi polynomials. In the fifth section \ref{sec:S1}, we use the same method and determine the exact solutions of the DKP oscillator for both natural and unnatural parities of spin 1 case. We will derive the solutions for dS models from the ones corresponding to AdS spaces in the penultimate section \ref{sec:dS} and finally, the concluding remarks come in the last section \ref{sec:CC}.

\section{Review of the deformed quantum mechanics relation}
\label{sec:RV}
In three--dimensional case, the deformed Heisenberg algebra leading to EUP of dS and AdS Spaces is defined by the following commutation relations \cite{Mignemi12, Stetsko15}:
\begin{equation}
\left[X_{i},X_{j}\right]=0\text{ , }\left[P_{i},P_{j}\right] =-i\hbar
\lambda \epsilon _{ijk}L_{k}\text{ , }\left[ X_{i},P_{j}\right] =i\hbar
\left( \delta _{ij}+\lambda X_{i}X_{j}\right)  \tag{1}  \label{eqt1}
\end{equation}
where $\lambda $ is a small parameter related to the deformation; it is positive for AdS case and negative for dS one. For example in the context of quantum gravity, this EUP parameter $\lambda$ is determined as the fundamental constant associated to the scale factor of the expanding universe and it is proportional to the cosmological constant $\Gamma=-3\lambda =-3a^{-2}$ where $a$ is the AdS radius \cite{Bolen05}.

$L_{k}$ are the component of the angular momentum expressed as follows:
\begin{equation}
L_{k}=\epsilon _{ijk}X_{i}P_{j}  \tag{2}  \label{eqt2}
\end{equation}
and it satisfies the usual algebra:
\begin{equation}
\left[ L_{i},P_{j}\right] =i\hbar \varepsilon _{ijk}P_{k}\text{ , }\left[
L_{i},X_{j}\right] =i\hbar \varepsilon _{ijk}X_{k}\text{ , }\left[
L_{i},L_{j}\right] =i\hbar \varepsilon _{ijk}L_{k}  \tag{3}  \label{eqt3}
\end{equation}
The deformed algebra of the AdS model from \ref{eqt1} is characterized by the presence of a nonzero minimum uncertainty in momentum and it gives rise to modified Heisenberg uncertainty relations:
\begin{equation}
\Delta X_{i}\Delta P_{i}\geq \frac{\hbar }{2}\left( 1+\lambda \left( \Delta
X_{i}\right) ^{2}\right)  \tag{4}  \label{eqt4}
\end{equation}
where we have chosen the states for which $\left\langle X_{i}\right\rangle =0$.

For simplicity, we assume isotropic uncertainties $X_{i}=X$, therefore, we arrive to a minimal uncertainty in momentum for the AdS model given by:
\begin{equation}
\left( \Delta P_{i}\right) _{\min}^{AdS}=\hbar \sqrt{\lambda }  \tag{5} \label{eqt5}
\end{equation}

\begin{figure}
\centering

\includegraphics[width=0.5\textwidth]{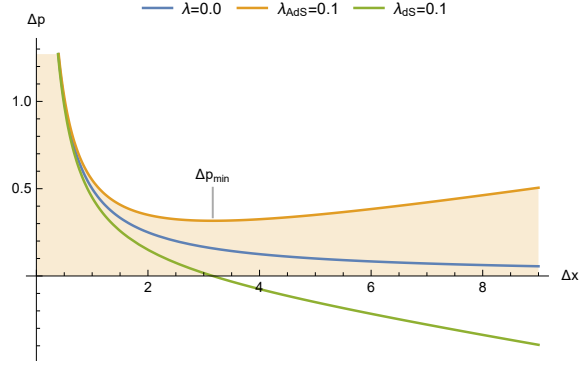}
\caption{$\Delta P$ versus $\Delta X$ in both AdS and dS spaces (forbidden regions for AdS case are colored)}
\label{fig1}
\end{figure}

This is shown in fig\ref{fig1} (we use the units $\hbar =c=m=1$ in all figures), where the usual Heisenberg uncertainty relation is plotted along with the modified relation found in \ref{eqt4} and where we have colored the forbidden region for the AdS case $\lambda \neq 0$ to show the limits on $\Delta P$ values.

For dS model, it suffices to invert the sign of $\lambda$ in \ref{eqt1} and \ref{eqt4} to write the corresponding relations. Of course there is no minimal uncertainty in momentum is this case as one can see in fig\ref{fig1}.

In the following sections, we will employ the noncommutative operators $X_{i}$ and $P_{i}$ satisfying the AdS algebra \ref{eqt1} which gives rise to rescaled uncertainty relation \ref{eqt4} in momentum space. In order to study the exact solutions of the deformed DKP oscillator equation in 3D with EUP, we represent these operators as functions of the usual operators $x_{i}$ and $p_{i}$, satisfying the ordinary canonical commutation relations in position space; this is done with the following transformations:
\begin{align}
X_{i}& =\frac{x_{i}}{\sqrt{1-\lambda r^{2}}}  \tag{6a}  \label{eqt6a} \\
P_{i}& =-i\hbar \sqrt{1-\lambda r^{2}}\partial _{x_{i}}  \tag{6b}
\label{eqt6b}
\end{align}
where the variable $r$ vary in the domain $\left] -1/\sqrt{\lambda },1/\sqrt{\lambda }\right[ $.

\section{Nikiforov--Uvarov method}
\label{sec:NU}
The method is based on the hypergeometric differential equation and its aim is to reduce second order differential equations to this type with an appropriate coordinate transformation $s=s\left(x\right)$:
\begin{equation}
\psi ^{\backprime \backprime }\left( s\right) +\frac{\tilde{\tau}\left(
s\right) }{\sigma \left( s\right) }\psi ^{\backprime }\left( s\right) +\frac{
\tilde{\sigma}\left( s\right) }{\sigma ^{2}\left( s\right) }\psi \left(
s\right) =0  \tag{7}  \label{eqt7}
\end{equation}
where $\sigma\left(s\right)$ and $\widetilde{\sigma}\left(s\right)$ are polynomials of degree two at most and $\widetilde{\tau}\left(s\right)$ is a first degree polynomial at most \cite{Nikiforov88, Egrifes99}. If we take the following factorization $\psi\left(s\right)=\phi\left(s\right)y\left(s\right)$ in \ref{eqt7} we get \cite{Nikiforov88}:
\begin{equation}
\sigma \left( s\right) y^{\backprime \backprime }\left( s\right) +\tau
\left( s\right) y^{\backprime }\left( s\right) +\Lambda y\left( s\right) =0
\tag{8}  \label{eqt8}
\end{equation}
where the polynomial factors are given by:
\begin{equation}
\pi \left( s\right) =\sigma \left( s\right) \frac{d}{ds}\left( \ln \phi
\left( s\right) \right) \text{\ and }\tau \left( s\right) =\tilde{\tau}
\left( s\right) +2\pi \left( s\right)  \tag{9}  \label{eqt9}
\end{equation}
and $\Lambda $ is defined with:
\begin{equation}
\Lambda _{n}+n\tau ^{\backprime }+\frac{n\left( n-1\right) \sigma
^{\backprime \backprime }}{2}=0\text{ , }n=0,1,2,...  \tag{10}  \label{eqt10}
\end{equation}
To find the energy eigenvalues, we first have to determine $\pi\left(s\right)$ and $\Lambda$ by defining $k=\Lambda -\pi^{\backprime}\left(s\right)$ and solving the resulting quadratic equation for $\pi\left(s\right)$; so we get $\pi\left( s\right)$ as a polynomial of $s$:
\begin{equation}
\pi \left( s\right) =\left( \frac{\sigma ^{\backprime }-\widetilde{\tau }}{2}
\right) \pm \sqrt{\left( \frac{\sigma ^{\backprime }-\tilde{\tau}}{2}\right)
^{2}-\tilde{\sigma}+\sigma k}  \tag{11}  \label{eqt11}
\end{equation}

The determination of $k$ is the key point in computing $\pi \left( s\right) $ and it is done by setting that the expression in the root must be a square of a polynomial; this gives a quadratic equation for $k$.

To determine the polynomial solutions $y_{n}\left( s\right) $, we use \ref{eqt9} and the Rodrigues relation:
\begin{equation}
y_{n}\left( s\right) =\frac{C_{n}}{\rho \left( s\right) }\frac{d^{n}}{ds^{n}}
\left[ \sigma ^{n}\left( s\right) \rho \left( s\right) \right]  \tag{12}
\label{eqt12}
\end{equation}
where $C_{n}$ is a normalizable constant and the weight function $\rho\left(s\right) $ satisfies the following relation:
\begin{equation}
\frac{d}{ds}\left[ \sigma \left( s\right) \rho \left( s\right) \right] =\tau
\left( s\right) \rho \left( s\right)  \tag{13}  \label{eqt13}
\end{equation}
This last equation refers to classical orthogonal polynomials and we write for $y_{n}\left( s\right)$:
\begin{equation}
\int_{a}^{b}y_{n}\left( s\right) y_{m}\left( s\right) \rho \left( s\right)
ds=0\text{ , }m\neq n  \tag{14}  \label{eqt14}
\end{equation}
This completely determines the solutions with the NU method.

\section{Spin 0 DKP Oscillator}
\label{sec:S0}
We start first with some useful formulas of the DKP equation then we apply them in our system.

The DKP equation describing a free scalar and vector boson \cite{Moussa18, Duffin38, Nedjadi941} is written as:
\begin{equation}
\left[ c\vec{\beta}\cdot\vec{p}+mc^{2}\right] \Psi =i\hbar \beta ^{0}\left(
\frac{\partial \Psi }{\partial t}\right)  \tag{15}  \label{eqt15}
\end{equation}
where $m$ is the mass and $\beta^{\mu}$ are the DKP matrices (with $\mu =0,1,2,3$); one can find their properties listed in several works and we cite for example \cite{Chetouani04, Merad07, Cardoso16, Montigny19}.

We write the DKP oscillator in 3D space by analogy with the Dirac oscillator \cite{Moshinsky89}, so we introduce the non--minimal substitution \cite{Nedjadi941}:
\begin{equation}
\vec{p}\rightarrow \vec{p}-im\omega \eta ^{0}\vec{r}  \tag{16}
\label{eqt16}
\end{equation}
where $\omega$ is the oscillator frequency and $\eta^{0}$ is a matrix defined by $\eta^{0}=2\left(\beta^{0}\right)^{2}-1$, with $\left(\eta^{0}\right)^{2}=1$. With this substitution, we obtain the new equation for the DKP oscillator:
\begin{equation}
\left[ c\vec{\beta}\cdot \left( \vec{p}-im\omega \eta ^{0}\vec{r}\right)
+mc^{2}\right] \Psi =i\hbar \beta ^{0}\left( \frac{\partial \Psi }{\partial t}\right)  \tag{17}  \label{eqt17}
\end{equation}
The non--commuting coordinate and momentum operators due to the presence of EUP are expressed in terms of the commuting operators as following:
\begin{equation}
\vec{r}\rightarrow \frac{\vec{r}}{\sqrt{1-\lambda r^{2}}}\text{ \ and \ }
\vec{p}\rightarrow \sqrt{1-\lambda r^{2}}\vec{p}  \tag{18}  \label{eqt18}
\end{equation}
and using $\Psi\left(\mathbf{r,}t\right)=e^{-iEt/\hbar}\tilde{\Psi}\left( \mathbf{r}\right)$ in \ref{eqt17}, we get the stationary DKP equation:
\begin{equation}
\left[ c\vec{\beta}\cdot \left( \sqrt{1-\lambda r^{2}}\vec{p}-im\omega \eta
^{0}\frac{\vec{r}}{\sqrt{1-\lambda r^{2}}}\right) +mc^{2}\right] \tilde{\Psi}
\left( \mathbf{r}\right) =E\beta ^{0}\tilde{\Psi}\left( \mathbf{r}\right)
\tag{19}  \label{eqt19}
\end{equation}

We note here that, because of the symmetry of the problem, the components of the wave function $\tilde{\Psi}\left( \mathbf{r}\right) $ are also eigenfunctions of both $J^{2}$ and $J_{z}$ with the eigenvalues $J\left(J+1\right)$ and $M$ respectively; here $\vec{J}$ represents the total angular momentum and it is defined as the sum of the orbital angular momentum $\vec{L}$ and the spin $\vec{S}$. This total angular momentum commutes with the external potential because this later in central and with $\beta ^{0}$ and so it is a constant of motion.

For spin $0$ particles, the wave function is a vector with five components. We will use the method of the spherical coordinates in momentum space \cite{Hill54, Edmonds57} to write the wave function because it is more convenient for the symmetry of the problem and because, as we will mention later, this method is applicable for spin 1 case too. In this formulation, the five components wave function of a scalar particle $\Psi_{JM}\left(\mathbf{r}\right)$ is given by \cite{Nedjadi941}:
\begin{equation}
\Psi _{JM}\left( \mathbf{r}\right) =\left(
\begin{array}{c}
F_{nJ}\left( r\right) Y_{JM}\left( \Omega \right) \\
G_{nJ}\left( r\right) Y_{JM}\left( \Omega \right) \\
i\sum\limits_{L}H_{nJL}\left( r\right) Y_{JL1}^{M}\left( \Omega \right)
\end{array}
\right)  \tag{20}  \label{eqt20}
\end{equation}
In this expression $F_{nJ}\left(r\right)$, $G_{nJ}\left(r\right)$ and $H_{nJL}\left(r\right)$ are the radial wave functions, $Y_{JM}\left(\Omega\right)$ are the usual (or scalar) spherical harmonics and $Y_{JL1}^{M}\left(\Omega\right)$ are the normalised vector spherical harmonics.

We insert this new form of $\Psi_{JM}\left(\mathbf{r}\right)$ into \ref{eqt19} and we use the properties of vector spherical harmonics in position space representation (\cite{Hill54, Edmonds57}) to keep only the radial functions in the equations; we obtain the following coupled system:
\begin{align}
H& =0  \tag{21a}  \label{eqt21a} \\
mc^{2}G& =EF  \tag{21b}  \label{eqt21b} \\
mc^{2}H_{-1}& =c\zeta _{J}\left[ \hbar \sqrt{1-\lambda r^{2}}\left( \frac{d}{
dr}+\frac{J+1}{r}\right) +\frac{m\omega r}{\sqrt{1-\lambda r^{2}}}\right] F
\tag{21c}  \label{eqt21c} \\
mc^{2}H_{+1}& =-c\xi _{J}\left[ \hbar \sqrt{1-\lambda r^{2}}\left( \frac{d}{
dr}-\frac{J}{r}\right) +\frac{m\omega r}{\sqrt{1-\lambda r^{2}}}\right] F
\tag{21d}  \label{eqt21d} \\
mc^{2}F-EG& =c\left[ -\xi _{J}\left( \hbar \sqrt{1-\lambda r^{2}}\left(
\frac{d}{dr}+\frac{J+2}{r}\right) -\frac{m\omega r}{\sqrt{1-\lambda r^{2}}}
\right) H_{+1}\right.  \notag \\
& \left. +\zeta _{J}\left( \hbar \sqrt{1-\lambda r^{2}}\left( \frac{d}{dr}-
\frac{J-1}{r}\right) -\frac{m\omega r}{\sqrt{1-\lambda r^{2}}}\right) H_{-1}
\right]  \tag{21e}  \label{eqt21e}
\end{align}
where we used the following parameters and notations:
\begin{gather}
\xi _{J}=\sqrt{\left( J+1\right) /\left( 2J+1\right) }\text{ , }\zeta _{J}=
\sqrt{J/\left( 2J+1\right) }  \notag \\
F_{nJ}\left( r\right) =F\text{ , }G_{nJ}\left( r\right) =G\text{ , }
H_{nJJ}\left( r\right) =H\text{ , }H_{nJJ\pm 1}\left( r\right) =H_{\pm 1}
\tag{22}  \label{eqt22}
\end{gather}
If we insert equations \ref{eqt21a}, \ref{eqt21b}, \ref{eqt21c} and \ref{eqt21d} into \ref{eqt21e}, we remain with one equation for $F\left(r\right)$:
\begin{equation}
\left[ \left( \sqrt{1-\lambda r^{2}}\frac{d}{dr}\right) ^{2}+\frac{2\left(
1-\lambda r^{2}\right) }{r}\frac{d}{dr}-\frac{J\left( J+1\right) \left(
1-\lambda r^{2}\right) }{r^{2}}-\frac{\eta r^{2}}{\left( 1-\lambda
r^{2}\right) }+\varepsilon \right] F\left( r\right) =0  \tag{23}
\label{eqt23}
\end{equation}
with:
\begin{equation}
\eta =\frac{m\omega }{\hbar }\left( \frac{m\omega }{\hbar }-\lambda \right)
\text{ and }\varepsilon =\frac{E^{2}-m^{2}c^{4}}{\left( \hbar c\right) ^{2}}+
\frac{3m\omega }{\hbar }  \tag{24}  \label{eqt24}
\end{equation}
We transform \ref{eqt23} using the following transformations $F\left(r\right)=y^{\mu}g\left(y\right)$ and $y=\sqrt{1-\lambda r^{2}}$ to get:
\begin{equation}
\left[ \left( 1-y^{2}\right) \frac{d^{2}}{dy^{2}}+\left( \frac{2\mu }{y}
-\left( 2\mu +3\right) y\right) \frac{d}{dy}-\frac{J\left( J+1\right) y^{2}}{
1-y^{2}}+\frac{\varepsilon}{\lambda} -3\mu \right] g\left( y\right) =0  \tag{25} \label{eqt25}
\end{equation}
where $\mu $ verifies the relation $\eta-\lambda^{2}\mu\left(\mu-1\right)=0$. Solving this later gives us two solutions:
\begin{equation}
\mu _{1}=1-\frac{m\omega }{\lambda \hbar }\text{ and }\mu _{2}=\frac{m\omega
}{\lambda \hbar }  \tag{26}  \label{eqt26}
\end{equation}
The accepted value of $\mu $ in \ref{eqt26} is the second solution because, from the expression of $F\left( r\right) $, the function $g\left( y\right) $ should be nonsingular at $y=\pm 1$; so $\mu =\mu _{2}=m\omega /\lambda \hbar $.

In addition, we note that \ref{eqt25} possesses three singular points $y=0,\pm 1$ and to reduce it to a class of known differential equation with a polynomial solution, we use a new variable $s=2y^{2}-1$:
\begin{equation}
\left[ \left( 1-s^{2}\right) ^{2}\dfrac{d^{2}}{ds^{2}}+\left( 1-s^{2}\right)
\left( \left( \mu -1\right) -\left( \mu +2\right) s\right) \dfrac{d}{ds}
+a_{1}s^{2}+a_{2}s+a_{3}\right] g\left( s\right) =0  \tag{27}  \label{eqt27}
\end{equation}
The parameters $a_{1}$, $a_{2}$ and $a_{3}$ are defined by:
\begin{equation}
a_{1,3}=\frac{-1}{4}\left( J\left( J+1\right) \pm \frac{\varepsilon }{
\lambda }\mp 3\mu \right) \text{ and }a_{2}=\frac{-J\left( J+1\right) }{2}
\tag{28}  \label{eqt28}
\end{equation}
We see that equation \ref{eqt27} for $g\left( s\right) $ is similar to equation \ref{eqt7} for $\psi \left( s\right) $ and this enables us to use the NU method with the following expressions for the NU polynomials:
\begin{equation}
\sigma \left( s\right) =\left( 1-s^{2}\right) \text{ , }\tilde{\tau}\left(
s\right) =\left( \mu -1\right) -\left( \mu +2\right) s\text{ \ and }\tilde{
\sigma}\left( s\right) =a_{1}s^{2}+a_{2}s+a_{3}  \tag{29}  \label{eqt29}
\end{equation}
Substituting them into \ref{eqt11}, we obtain:
\begin{equation}
\pi \left( s\right) =\frac{\left( \mu s-\left( \mu -1\right) \right) }{2}\pm
\sqrt{\left( \frac{\mu ^{2}}{4}-a_{1}-k\right) s^{2}-\left( \frac{\mu \left(
\mu -1\right) }{4}+a_{2}\right) s+\frac{\left( \mu -1\right) ^{2}}{4}-a_{3}+k
}  \tag{30}  \label{eqt30}
\end{equation}
The parameter $k$ is determined as mentioned in the precedent section and we get two values:
\begin{equation}
k_{1}=\frac{1}{4}\left[ \frac{\varepsilon }{\lambda }-3\mu +\left( 2\mu
-1\right) \left( J+1\right) \right] \text{ and }k_{2}=\frac{1}{4}\left[ \frac{
\varepsilon }{\lambda }-3\mu -\left( 2\mu -1\right) J\right]  \tag{31} \label{eqt31}
\end{equation}

For $\pi\left(s\right)$, we obtain the following possible solutions:
\begin{equation}
\pi \left( s\right) =\left\{
\begin{array}{c}
\pi _{1}=\frac{1}{2}\left( \left( 2\mu -J-1\right) s-\left( 2\mu +J-1\right)
\right) \\
\pi _{2}=\frac{1}{2}\left( J+1\right) \left( s+1\right) \\
\pi _{3}=\frac{1}{2}\left( \left( 2\mu +J\right) s-\left( 2\mu -J-2\right)
\right) \\
\pi _{4}=-\frac{1}{2}J\left( s+1\right)
\end{array}
\right.  \tag{32}  \label{eqt32}
\end{equation}
where $\pi_{1}$ and $\pi_{2}$ are related to $k_{1}$ while $\pi_{3}$ and $\pi_{4}$ are linked to $k_{2}$. The correct solution is $\pi_{4}$, so:
\begin{equation}
\tau \left( s\right) =-\left( \mu +J+2\right) s+\left( \mu -J-1\right)
\tag{33}  \label{eqt33}
\end{equation}
From \ref{eqt10}, we obtain:
\begin{equation}
\Lambda =k_{2}-\frac{J}{2}=n\left( n+\mu +J+1\right) \text{ , }n=0,1,2,...
\tag{34}  \label{eqt34}
\end{equation}
Hence, the energy eigenvalues are found as ($N=2n+J$ is the principal quantum number):
\begin{equation}
E_{N,J}^{2}=m^{2}c^{4}+2\hbar m\omega c^{2}N+\lambda \hbar ^{2}c^{2}\left[
\left( N+1\right) ^{2}-J\left( J+1\right) -1\right]  \tag{35}  \label{eqt35}
\end{equation}

We remark that the above expression of the energies contains the usual 3D DKP oscillator term and an additional correction term depending on the deformation; the later is linearly proportional to the deformation parameter $\lambda $. Here it should be noted that the presence of a correction term proportional to $N^{2}$ indicates the appearance of a hard confinement due to the deformation. This is equivalent to the energy of a spinless relativistic quantum particle in a square well potential whose boundaries are placed at $\pm
\pi /2\sqrt{\lambda }$. The second term in the $\lambda$ correction is proportional to $J\left(J+1\right)$, so it appears as some kind of rotational energy and it removes the degeneracy of the usual oscillator spectrum according to this number $J$.

We note here that the spectral corrections due to the EUP are qualitatively different to those associated to the GUP \cite{Falek09, Falek10}. One can also recover the usual spectrum energy using the limit $\lambda \rightarrow 0$ and it coincides with that of the ordinary spinless 3D DKP oscillator \cite{Nedjadi941}.

\begin{figure}
\centering
\includegraphics[width=0.5\textwidth]{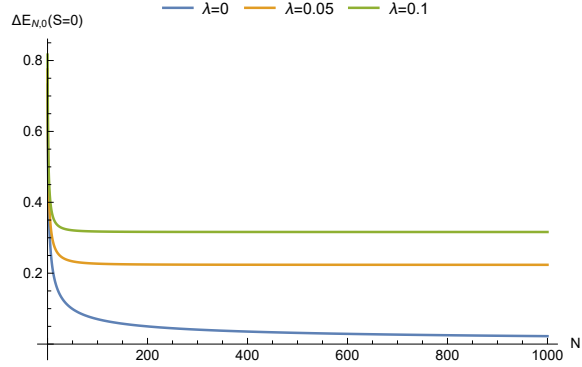}
\caption{Energy spacing $\Delta E=E(N+1)-E(N)$ with and without deformation}
\label{fig2}
\end{figure}

Another interesting characteristic of this spectrum is the limit of the energy levels spacing:
\begin{equation}
\lim_{N\rightarrow \infty }\left\vert \Delta E_{N,J}\right\vert =\hbar c\sqrt{\lambda }  \tag{36}  \label{eqt36}
\end{equation}
We see from \ref{eqt35} and fig\ref{fig2} that this difference becomes constant for large values of $N$ and the spectrum remains bounded even at the limit $N\rightarrow \infty$. In the ordinary case when the parameter $\lambda$ vanishes, this energy spacing tends to zero for large $N$ and the spectrum becomes almost continuous.

To obtain an upper bound on the deformation parameter, we expand \ref{eqt35} to the first order in $\lambda $:
\begin{equation}
E_{N,0}=\sqrt{m^{2}c^{4}+2\hbar m\omega c^{2}N}+\frac{\hbar ^{2}c^{2}N\left(
N+2\right) \lambda }{2\sqrt{m^{2}c^{4}+2\hbar m\omega c^{2}N}}  \tag{37} \label{eqt37}
\end{equation}
The deviation of the $N$-th energy level caused by the modified commutation relations \ref{eqt1} is:
\begin{equation}
\frac{\Delta E_{N,0}}{\hbar \omega }=\frac{\lambda \hbar ^{2}N\left(
N+2\right) }{2m\hbar \omega \sqrt{1+2\hbar \omega m^{-1}c^{-2}N}}  \tag{38} \label{eqt38}
\end{equation}
We use experimental results of the cyclotron motion of an electron in a Penning trap \cite{Brown86, Mittleman99} where $\omega_{c}=eB/m_{e}$ is its cyclotron frequency when it is trapped in a magnetic field $B$. Therefore, for a magnetic field of strength $B=6T$ (in IS units), we have $m_{e}\hbar \omega _{c}=e\hbar B=10^{-52}kg^{2}m^{2}s^{-2}$. If we assume that at the level $n=10^{10}$, only a deviation of the $\hbar\omega_{c}$ scale can be detected and by taking $\Delta E_{n}<\hbar\omega_{c}$ (no perturbation is observed for the $n$--th energy level) \cite{Chang02}, we get the following upper bound for the minimal uncertainty in momentum:
\begin{equation}
\Delta P_{\min }=\hbar \sqrt{\lambda }<3.25\times 10^{-36}kgms^{-1}  \tag{39} \label{eqt39}
\end{equation}
Now we focus on the corresponding eigenfunctions. Taking the expression of $\pi _{4}\left( s\right) $ from \ref{eqt32}, the $\phi \left( s\right) $ part is defined from \ref{eqt9} as $\phi \left( s\right) =\left( 1-s\right)^{J/2}$ and according to the form of $\sigma \left( s\right) $ in \ref{eqt29}, the $y\left( s\right) $ part comes from the Rodrigues relation \ref{eqt12}:
\begin{equation}
y_{n}\left( s\right) =\frac{C_{n}}{\rho \left( s\right) }\frac{d^{n}}{ds^{n}}
\left[ \left( 1-s^{2}\right) ^{n}\rho \left( s\right) \right]  \tag{40} \label{eqt40}
\end{equation}
where the weight function $\rho \left( s\right) $ is determined from the expressions of $\tau\left(s\right)$ and $\sigma\left(s\right)$:
\begin{equation}
\rho \left( s\right) =\left( 1+s\right) ^{\mu -1/2}\left( 1-s\right) ^{J+1/2}
\tag{41}  \label{eqt41}
\end{equation}
The relation \ref{eqt40} stands for the Jacobi polynomials, so we get:
\begin{equation}
y_{n}\left( s\right) \equiv P_{n}^{\left( J+1/2,\mu -1/2\right) }\left(
s\right)  \tag{42}  \label{eqt42}
\end{equation}
Hence, $g(s)$ is written from its definition as follows:
\begin{equation}
g(s)=\phi \left( s\right) y_{n}\left( s\right) =C_{n}\left( 1-s\right)
^{J/2}P_{n}^{\left( J+1/2,\mu -1/2\right) }\left( s\right)  \tag{43} \label{eqt43}
\end{equation}
This allows us to write the general form of the component $F$ in terms of the variable $r$ as follows:
\begin{equation}
F_{n}\left( r\right) =C_{n}\left( 1-\lambda r^{2}\right) ^{\frac{\mu }{2}
}\left( 2\lambda r^{2}\right) ^{\frac{J}{2}}P_{n}^{\left( J+1/2,\mu
-1/2\right) }\left( 1-2\lambda r^{2}\right)  \tag{44}  \label{eqt44}
\end{equation}
where $C_{n}$ is the normalization constant.

Now to recover $\Psi_{nJ}\left(\mathbf{r}\right)$, we use the following property of Jacobi polynomials \cite{Gradshteyn80}:
\begin{equation}
\frac{dP_{n}^{\left( a,b\right) }\left( y\right) }{dy}=\frac{1}{2}\left(
n+a+b+1\right) P_{n-1}^{\left( a+1,b+1\right) }\left( y\right)  \tag{45} \label{eqt45}
\end{equation}
to obtain its other components from equations 25 and then its final form from \ref{eqt20}:
\begin{align}
H &=0\text{ , }G=C_{n}\frac{E}{mc^{2}}\left( 1-\lambda r^{2}\right) ^{\frac{
\mu }{2}}\left( 2\lambda r^{2}\right) ^{\frac{J}{2}}P_{n}^{\left( J+1/2,\mu
-1/2\right) }\left( 1-2\lambda r^{2}\right)  \notag \\
H_{+1} &=-iC_{n}\xi _{J}\frac{\hbar }{mc}\left( 1-\lambda r^{2}\right) ^{
\frac{\mu +1}{2}}\left( 2\lambda r^{2}\right) ^{\frac{J}{2}}\left( \frac{
m\omega }{\hbar }+\left( n+J+1\right) \lambda \right) rP_{n-1}^{\left(
J+3/2,\mu +1/2\right) } \tag{46}  \label{eqt46} \\
H_{-1} &=iC_{n}\zeta _{J}\frac{\hbar }{mc}\left( 1-\lambda r^{2}\right) ^{
\frac{\mu +1}{2}}\left( 2\lambda r^{2}\right) ^{\frac{J}{2}}\left[ \frac{
\left( 2J+1\right) }{r}P_{n}^{\left( J+\frac{1}{2},\mu -\frac{1}{2}\right) }
\right] -\frac{\zeta _{J}}{\xi _{J}}H_{+1}  \notag
\end{align}

\section{Spin 1 DKP Oscillator}
\label{sec:S1}
In this case, the wave function has ten components and we have to use the spherical spatial form of the ten components wave function $\Psi \left( \mathbf{r}\right) $ in the momentum space, otherwise we will not be able to decouple the system; so we write:
\begin{equation}
\Psi _{JM}\left( \mathbf{r}\right) =\left(
\begin{array}{c}
i\phi _{nJ}\left( r\right) Y_{JM}\left( \Omega \right) \\
\sum\limits_{L}F_{nJL}\left( r\right) Y_{JL1}^{M}\left( \Omega \right) \\
\sum\limits_{L}G_{nJL}\left( r\right) Y_{JL1}^{M}\left( \Omega \right) \\
\sum\limits_{L}H_{nJL}\left( r\right) Y_{JL1}^{M}\left( \Omega \right)
\end{array}
\right)  \tag{47}  \label{eqt47}
\end{equation}
Here, $\Psi _{JM}$, $F_{nJ}\left(r\right)$, $G_{nJ}\left(r\right)$ and $H_{nJL}\left(r\right)$ are the radial wave functions, while $Y_{JM}\left(\Omega\right)$ are the scalar spherical harmonics and $Y_{JL1}^{M}\left(\Omega\right)$ are the normalised vector spherical harmonics.

Putting this form of $\Psi _{JM}\left( \mathbf{r}\right) $ into \ref{eqt19} leads to ten coupled differential radial equations which can be reduced to two classes associated with the two parities $\left( -1\right) ^{J}$ and $\left(-1\right) ^{J+1}$ \cite{Nedjadi941, Kulikov05}.

\subsection{Natural Parity States}

Here the parity is $\left( -1\right) ^{J}$ and the relevant radial differential equations are:
\begin{align}
mc^{2}G_{0} &=EF_{0} \tag{48a} \label{eqt48a} \\
mc^{2}H_{-1} &=-c\xi _{J}\left[ \hbar \sqrt{1-\lambda r^{2}}\left( \frac{d}{
dr}+\frac{J+1}{r}\right) +\frac{m\omega r}{\sqrt{1-\lambda r^{2}}}\right]
F_{0} \tag{48b} \label{eqt48b} \\
mc^{2}H_{+1} &=-c\zeta _{J}\left[ \hbar \sqrt{1-\lambda r^{2}}\left( \frac{d
}{dr}-\frac{J}{r}\right) +\frac{m\omega r}{\sqrt{1-\lambda r^{2}}}\right]
F_{0} \tag{48c} \label{eqt48c} \\
mc^{2}F_{0}-EG_{0} &=-c\xi _{J}\left[ \hbar \sqrt{1-\lambda r^{2}}\left(
\frac{d}{dr}-\frac{J-1}{r}\right) -\frac{m\omega r}{\sqrt{1-\lambda r^{2}}}
\right] H_{-1}  \notag \\
&-c\zeta _{J}\left[ \hbar \sqrt{1-\lambda r^{2}}\left( \frac{d}{dr}+\frac{
J+2}{r}\right) -\frac{m\omega r}{\sqrt{1-\lambda r^{2}}}\right] H_{+1}
\tag{48d} \label{eqt48d}
\end{align}
with the following notations $R_{nJJ}\left(r\right)=R_{0}$, $R_{nJJ\pm 1}\left(r\right)=R_{\pm}$ and $R$ designates $F$, $G$ and $H$.

We eliminate $G_{0}$, $H_{+1}$ and $H_{-1}$ in favour of $F_{0}$ to obtain the following differential equation:
\begin{equation}
\left[ \left( \sqrt{1-\lambda r^{2}}\frac{d}{dr}\right) ^{2}+\frac{2\left(
1-\lambda r^{2}\right) }{r}\frac{d}{dr}-\frac{J\left( J+1\right) \left(
1-\lambda r^{2}\right) }{r^{2}}-\frac{\eta r^{2}}{\left( 1-\lambda
r^{2}\right) }+\varepsilon ^{\backprime }\right] F_{0}=0  \tag{49} \label{eqt49}
\end{equation}
where:
\begin{equation}
\eta =\frac{m\omega }{\hbar }\left( \frac{m\omega }{\hbar }-\lambda \right)
\text{ and }\varepsilon ^{\backprime }=\frac{E^{2}-m^{2}c^{4}}{\left( \hbar
c\right) ^{2}}+\frac{m\omega }{\hbar }-\lambda  \tag{50}  \label{eqt50}
\end{equation}

We note that equation \ref{eqt49} is similar to \ref{eqt23} and so we can solve it exactly in the same manner to obtain the $F_{0}$ function and then the other components from the equation system $48$; this enables us to get the complete solutions for the eigenfunctions of natural parity states:
\begin{align}
F_{0} &=C_{n}^{\backprime }\left( 1-\lambda r^{2}\right) ^{\frac{\mu }{2}
}\left( 2\lambda r^{2}\right) ^{\frac{J}{2}}P_{n}^{\left( J+1/2,\mu
-1/2\right) }\left( 1-2\lambda r^{2}\right) \tag{51a} \label{eqt51a} \\
G_{0} &=C_{n}^{\backprime }\frac{E}{mc^{2}}\left( 1-\lambda r^{2}\right) ^{
\frac{\mu }{2}}\left( 2\lambda r^{2}\right) ^{\frac{J}{2}}P_{n}^{\left(
J+1/2,\mu -1/2\right) }\left( 1-2\lambda r^{2}\right) \tag{51b} \label{eqt51b} \\
H_{+1} &=C_{n}^{\backprime }\zeta _{J}\frac{\hbar }{mc}\left( 1-\lambda
r^{2}\right) ^{\frac{\mu +1}{2}}\left( 2\lambda r^{2}\right) ^{\frac{J}{2}}
\left[ \left( \frac{m\omega }{\hbar }+\left( n+J+1\right) \lambda \right)
rP_{n-1}^{\left( J+3/2,\mu +1/2\right) }\right] \tag{51c} \label{eqt51c} \\
H_{-1} &=-C_{n}^{\backprime }\xi _{J}\frac{\hbar }{mc}\left( 1-\lambda
r^{2}\right) ^{\frac{\mu +1}{2}}\left( 2\lambda r^{2}\right) ^{\frac{J}{2}}
\left[ \frac{\left( 2J+1\right) }{r}P_{n}^{\left( J+1/2,\mu -1/2\right) }
\right] +\frac{\xi _{J}}{\zeta _{J}}H_{+1} \tag{51d} \label{eqt51d}
\end{align}
where $C_{n}^{\backprime }$ is a normalization constant and $\mu =m\omega /\lambda \hbar $ from \ref{eqt26}.

\begin{figure}
\centering
\includegraphics[width=0.5\textwidth]{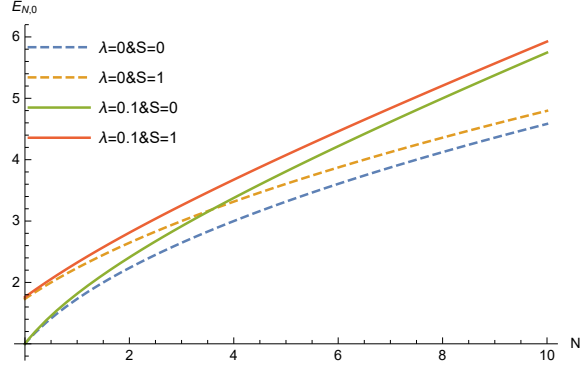}
\caption{$E_{N,0}$ for both spin 0 case and spin 1 natural states}
\label{fig3}
\end{figure}

The correspondent energy spectrum is given by:
\begin{equation}
E_{N,J}^{2}=m^{2}c^{4}+2\hbar m\omega c^{2}\left( N+1\right) +\lambda \hbar
^{2}c^{2}\left[ \left( N+1\right) ^{2}-J\left( J+1\right) \right]  \tag{52} \label{eqt52}
\end{equation}
We can easily compare this spectrum with the one corresponding to the spin 0 case \ref{eqt35}. We observe the same effects due to deformation, a strong confinement term and a rotational energy term which removes the $J$ degeneracy. Of course, the similarity is valid for the non--relativistic limits too:
\begin{align}
E_{NJ}^{nr}(S=1) &=\hbar \omega \left( N+1\right) +\frac{\lambda \hbar ^{2}}{2m}
\left[ \left( N+1\right) ^{2}-J\left( J+1\right) \right]  \tag{53} \label{eqt53} \\
E_{NJ}^{nr}(S=0) &=\hbar \omega N +\frac{\lambda \hbar ^{2}}{2m}\left[ \left(
N+1\right) ^{2}-J\left( J+1\right) -1\right]  \tag{54}  \label{eqt54}
\end{align}
The corrections have the same behavior in both cases; they increase with $N$ and decrease with $J$. We give an example of this similarity in the behaviors in fig\ref{fig3}, by representing the dependence with the values of $N$ of the relativistic energies for spin 0 particle (from \ref{eqt39}) and also for natural parity of spin 1 particle (from \ref{eqt52}) for the $J=0$ states.

\subsection{Unnatural Parity States}

The parity in this case is $\left( -1\right) ^{J+1}$ and the relevant system for radial functions is:
\begin{align}
mc^{2}F_{+1}-EG_{+1}& =-c\zeta _{J}\left[ \hbar \sqrt{1-\lambda r^{2}}\left(
\frac{d}{dr}-\frac{J}{r}\right) -\frac{m\omega r}{\sqrt{1-\lambda r^{2}}}
\right] H_{0}  \tag{55a}  \label{eqt55a} \\
mc^{2}F_{-1}-EG_{-1}& =-c\xi _{J}\left[ \hbar \sqrt{1-\lambda r^{2}}\left(
\frac{d}{dr}+\frac{J+1}{r}\right) -\frac{m\omega r}{\sqrt{1-\lambda r^{2}}}
\right] H_{0}  \tag{55b}  \label{eqt55b} \\
mc^{2}H_{0}& =-c\zeta _{J}\left[ \hbar \sqrt{1-\lambda r^{2}}\left( \frac{d}{
dr}+\frac{J+2}{r}\right) +\frac{m\omega r}{\sqrt{1-\lambda r^{2}}}\right]
F_{+1}  \notag \\
& -c\xi _{J}\left[ \hbar \sqrt{1-\lambda r^{2}}\left( \frac{d}{dr}-\frac{J-1
}{r}\right) +\frac{m\omega r}{\sqrt{1-\lambda r^{2}}}\right] F_{-1}
\tag{55c}  \label{eqt55c} \\
mc^{2}G_{+1}-EF_{+1}& =-c\xi _{J}\left[ \hbar \sqrt{1-\lambda r^{2}}\left(
\frac{d}{dr}-\frac{J}{r}\right) -\frac{m\omega r}{\sqrt{1-\lambda r^{2}}}
\right] \phi  \tag{65d}  \label{eqt65d} \\
mc^{2}G_{-1}-EF_{-1}& =c\zeta _{J}\left[ \hbar \sqrt{1-\lambda r^{2}}\left(
\frac{d}{dr}+\frac{J+1}{r}\right) -\frac{m\omega r}{\sqrt{1-\lambda r^{2}}}
\right] \phi  \tag{55e}  \label{eqt55e} \\
mc^{2}\phi & =-c\xi _{J}\left[ \hbar \sqrt{1-\lambda r^{2}}\left( \frac{d}{dr
}+\frac{J+2}{r}\right) +\frac{m\omega r}{\sqrt{1-\lambda r^{2}}}\right]
G_{+1}  \notag \\
& +c\zeta _{J}\left[ \hbar \sqrt{1-\lambda r^{2}}\left( \frac{d}{dr}-\frac{
J-1}{r}\right) +\frac{m\omega r}{\sqrt{1-\lambda r^{2}}}\right] G_{-1}
\tag{55f}  \label{eqt55f}
\end{align}
and we adopt the same notations as those already used for the functions in the system corresponding to natural parity states: $R_{nJJ}\left(r\right)=R_{0}$, $R_{nJJ\pm 1}\left(r\right)=R_{\pm}$ and $R$ designates $F$, $G$ and $H$.

To solve this system, we start by simplifying its writing to this compact shape:
\begin{align}
\left(
\begin{array}{c}
F_{+1} \\
G_{+1}
\end{array}
\right) & =\frac{\hbar }{\epsilon }\left( \sqrt{1-\lambda r^{2}}\left( \frac{
d}{dr}-\frac{J}{r}\right) -\frac{m\omega \hbar ^{-1}r}{\sqrt{1-\lambda r^{2}}
}\right) \left(
\begin{array}{cc}
\xi _{J}E & \zeta _{J}mc^{2} \\
\xi _{J}mc^{2} & \zeta _{J}E
\end{array}
\right) \left(
\begin{array}{c}
\phi \\
H_{0}
\end{array}
\right)  \tag{56a}  \label{eqt56a} \\
\left(
\begin{array}{c}
F_{-1} \\
G_{-1}
\end{array}
\right) & =\frac{\hbar }{\epsilon }\left( \sqrt{1-\lambda r^{2}}\left( \frac{
d}{dr}+\frac{J+1}{r}\right) -\frac{m\omega \hbar ^{-1}r}{\sqrt{1-\lambda
r^{2}}}\right) \left(
\begin{array}{cc}
-\zeta _{J}E & \xi _{J}mc^{2} \\
-\zeta _{J}mc^{2} & \xi _{J}E
\end{array}
\right) \left(
\begin{array}{c}
\phi \\
H_{0}
\end{array}
\right)  \tag{56b}  \label{eqt56b}
\end{align}
and:
\begin{align}
2\sqrt{J\left( J+1\right) }WEH_{0}& =\left[ \left( \sqrt{1-\lambda r^{2}}
\frac{d}{dr}\right) ^{2}+\frac{2\left( 1-\lambda r^{2}\right) }{r}\frac{d}{dr
}-\frac{J\left( J+1\right) \left( 1-\lambda r^{2}\right) }{r^{2}}\right.
\notag \\
& \left. -\frac{\eta r^{2}}{\left( 1-\lambda r^{2}\right) }+\frac{\epsilon c
}{\left( \hbar m\omega c\right) ^{2}}-\frac{3m\omega }{\hbar }\right] \phi
\tag{57a}  \label{eqt57a} \\
2\sqrt{J\left( J+1\right) }WE\phi & =\left[ \left( \sqrt{1-\lambda r^{2}}
\frac{d}{dr}\right) ^{2}+\frac{2\left( 1-\lambda r^{2}\right) }{r}\frac{d}{dr
}-\frac{J\left( J+1\right) \left( 1-\lambda r^{2}\right) }{r^{2}}\right.
\notag \\
& \left. -\frac{\eta r^{2}}{\left( 1-\lambda r^{2}\right) }+\frac{\epsilon c
}{\left( \hbar m\omega c\right) ^{2}}-\frac{m\omega }{\hbar }-\lambda \right]
H_{0}  \tag{57b}  \label{eqt57b}
\end{align}
Here $\epsilon=c^{-1}\left(E^{2}-m^{2}c^{4}\right)$ and $W=\left(1-\lambda\hbar/2m\omega\right)\omega /\hbar c^{2}$.

Now, according to the procedure of diagonalization \cite{Nedjadi941, Nedjadi942}:
\begin{equation}
\left(
\begin{array}{c}
\phi \\
H_{0}
\end{array}
\right) =\frac{1}{\sqrt{2\kappa \left( \kappa +1\right) }}\left(
\begin{array}{cc}
\left( 1+\kappa \right) & k \\
k & -\left( 1+\kappa \right)
\end{array}
\right) \left(
\begin{array}{c}
R_{+} \\
R_{-}
\end{array}
\right)  \tag{58}  \label{eqt58}
\end{equation}
with $k=2\sqrt{J\left( J+1\right) }E/mc^{2}$ and $\kappa =\sqrt{1+k^{2}}$, we can decuple \ref{eqt57a} and \ref{eqt57b} as follows:
\begin{equation}
\left[ \left( \sqrt{1-\lambda r^{2}}\frac{d}{dr}\right) ^{2}+\frac{2\left(
1-\lambda r^{2}\right) }{r}\frac{d}{dr}-\frac{J\left( J+1\right) \left(
1-\lambda r^{2}\right) }{r^{2}}-\frac{\eta r^{2}}{\left( 1-\lambda
r^{2}\right) }+\varepsilon _{+}\right] R_{+}=0  \tag{59}  \label{eqt59}
\end{equation}
\begin{equation}
\left[ \left( \sqrt{1-\lambda r^{2}}\frac{d}{dr}\right) ^{2}+\frac{2\left(
1-\lambda r^{2}\right) }{r}\frac{d}{dr}-\frac{J\left( J+1\right) \left(
1-\lambda r^{2}\right) }{r^{2}}-\frac{\eta r^{2}}{\left( 1-\lambda
r^{2}\right) }+\varepsilon _{-}\right] R_{-}=0  \tag{60}  \label{eqt60}
\end{equation}
where:
\begin{equation}
\varepsilon _{\pm }=\frac{E^{2}-m^{2}c^{4}}{\hbar ^{2}c^{2}}+\frac{\omega }{
\hbar c^{2}}\left( 1-\frac{\lambda \hbar }{2m\omega }\right) \left(
mc^{2}\mp \sqrt{m^{2}c^{4}+4J\left( J+1\right) E^{2}}\right) -\frac{3m\omega
}{\hbar }  \tag{61}  \label{eqt61}
\end{equation}

We remark that equations \ref{eqt59} and \ref{eqt60} are exactly the same as equation \ref{eqt23} corresponding to the spin 0 case and also to equation \ref{eqt49} of natural parity states; so we solve them in the same manner. The relativistic spinor wave functions $R_{\pm }$ are exactly the same as $F(r)$ in \ref{eqt44} and they are given using the Jacobi polynomials as:

\begin{equation}
R_{\pm }\left( r\right) =C_{\pm }\left( 1-\lambda r^{2}\right) ^{\frac{\mu }{
2}}\left( 2\lambda r^{2}\right) ^{\frac{J}{2}}P_{n}^{\left( J+1/2,\mu
-1/2\right) }\left( 1-2\lambda r^{2}\right)  \tag{62}  \label{eqt62}
\end{equation}
with $\mu=m\omega/\lambda\hbar$ and where $C_{\pm}$ are the normalization constants.

The determination of $R_{\pm }$ gives us both $\phi$ and $H_{0}$ from \ref{eqt58} and then, using \ref{eqt56a} and \ref{eqt56b}, the remaining components of the wave function $\Psi\left(\mathbf{r}\right)$ associated with the unnatural parity states $\left( -1\right) ^{J+1}$:
\begin{multline}
\left(
\begin{array}{c}
\phi \\
H_{0} \\
F_{+} \\
G_{+} \\
F_{-} \\
G_{-}
\end{array}
\right) =\frac{\left( 1-\lambda r^{2}\right) ^{\frac{\mu }{2}}\left(
2\lambda r^{2}\right) ^{\frac{J}{2}}}{\epsilon \sqrt{2\kappa \left( \kappa
+1\right) }}\left\{ \left[ C_{+}\left(
\begin{array}{c}
\left( 1+\kappa \right) \epsilon \\
k\epsilon \\
\alpha _{+}\Gamma _{1} \\
\beta _{+}\Gamma _{1} \\
\gamma _{+}\Gamma _{2} \\
\delta _{+}\Gamma _{2}
\end{array}
\right) +C_{-}\left(
\begin{array}{c}
k\epsilon \\
-\left( 1+\kappa \right) \epsilon \\
\delta _{-}\Gamma _{1} \\
\gamma _{-}\Gamma _{1} \\
\beta _{-}\Gamma _{2} \\
\alpha _{-}\Gamma _{2}
\end{array}
\right) \right] P_{n}^{\left( J+\frac{1}{2},\mu -\frac{1}{2}\right) }\right.
\notag \\
\left. -\left[ C_{+}\left(
\begin{array}{c}
0 \\
0 \\
\alpha _{+} \\
\beta _{+} \\
\gamma _{+} \\
\delta _{+}
\end{array}
\right) +C_{-}\left(
\begin{array}{c}
0 \\
0 \\
\delta _{-} \\
\gamma _{-} \\
\beta _{-} \\
\alpha _{-}
\end{array}
\right) \right] \Gamma _{3}P_{n-1}^{\left( J+\frac{3}{2},\mu +\frac{1}{2}
\right) }\right\} \left( 1-2\lambda r^{2}\right)  \tag{63}  \label{eqt63}
\end{multline}
with the following abbreviated notations:
\begin{align*}
\alpha _{\pm }& =\zeta _{J}mc^{2}k\pm \xi _{J}E\left( \kappa +1\right) \text{
, }\beta _{\pm }=-\zeta _{J}Ek\pm \xi _{J}mc^{2}\left( \kappa +1\right) \\
\gamma _{\pm }& =\xi _{J}mc^{2}k\pm \zeta _{J}E\left( \kappa +1\right) \text{
, }\delta _{\pm }=\xi _{J}Ek\pm \zeta _{J}mc^{2}\left( \kappa +1\right) \\
\Gamma _{1}\left( r\right) & =\frac{-2m\omega r}{\sqrt{1-\lambda r^{2}}}
\text{ , }\Gamma _{2}\left( r\right) =\frac{\hbar \left( 2J+1\right) \sqrt{
1-\lambda r^{2}}}{r}+\Gamma _{1}\left( r\right) \text{ , }\Gamma _{3}\left(
r\right) =\lambda r\sqrt{1-\lambda r^{2}}\left( n+\mu +J+1\right)
\end{align*}

We also derive the correspondent equations for the relativistic energies $E_{+}$and $E_{-}$ related to the spinors $R_{+}$ and $R_{-}$ respectively and this for all values of $\lambda $:
\begin{multline}
\frac{E_{\pm }^{2}-m^{2}c^{4}}{\hbar \omega }\mp \left( 1-\frac{\lambda
\hbar }{2m\omega }\right) \sqrt{m^{2}c^{4}+4J\left( J+1\right) E_{\pm }^{2}}=
\notag \\
mc^{2}\left( 2N+5\right) +\frac{\lambda \hbar c^{2}}{\omega }\left[ \left(
N+1\right) ^{2}-J\left( J+1\right) -\frac{1}{2}\right]  \tag{64} \label{eqt64}
\end{multline}

The exact solutions of these eigenvalue equations take the forms:
\begin{multline}
E_{\pm }^{2}=m^{2}c^{4}+2\hbar m\omega c^{2}\left( N+\frac{5}{2}\right)
+2\hbar ^{2}\omega ^{2}\left( 1-\frac{\lambda \hbar }{2m\omega }\right)
^{2}J\left( J+1\right) +  \notag \\
\lambda \hbar ^{2}c^{2}\left( \left( N+1\right) ^{2}-J\left( J+1\right) -
\frac{1}{2}\right) \pm \Delta  \tag{65}  \label{eqt65}
\end{multline}
where $\Delta$ is given by ($b_{0}=\left(2J+1\right)^{2}$ and $b_{1}=4J\left(J+1\right)$):
\begin{multline}
\Delta =\hbar m\omega c^{2}\left( 1-\frac{\lambda \hbar }{2m\omega }\right)
\left( 2J+1\right) \left[ 1+\frac{b_{1}^{2}}{4b_{0}}\left( \frac{\hbar
\omega }{mc^{2}}\right) ^{2}\left( 1-\frac{\lambda \hbar }{2m\omega }\right)
^{2}\right.  \notag \\
\left. +\frac{2b_{1}}{b_{0}}\frac{\hbar \omega }{mc^{2}}\left( N+\frac{5}{2}+
\frac{\lambda \hbar }{2m\omega }\left( \left( N+1\right) ^{2}-J\left(
J+1\right) -\frac{1}{2}\right) \right) \right] ^{1/2}  \tag{66} \label{eqt66}
\end{multline}

The limit $\lambda\rightarrow 0$ gives the ordinary spectrum and it coincides exactly with the one found in \cite{Nedjadi941}.

The non--relativistic spectrum is obtained using the usual approximation mentioned above:
\begin{multline}
E_{\pm }^{nr}=\hbar \omega \left( N+\frac{5}{2}\right) +\frac{\lambda \hbar
^{2}}{2m}\left( \left( N+1\right) ^{2}-J\left( J+1\right) -\frac{1}{2}
\right)  \notag \\
 +\frac{\hbar ^{2}\omega ^{2}}{mc^{2}}\left( 1-\frac{\lambda \hbar }{
2m\omega }\right) ^{2}J\left( J+1\right) \pm \frac{\Delta }{2mc^{2}}
\tag{67}  \label{eqt67}
\end{multline}

Comparing this formula with those of the natural spin 1 case \ref{eqt53} and of the spin 0 case \ref{eqt54}, we see that this spectrum differs by the last two terms in \ref{eqt67} and thus its dependence on the deformation is more pronounced. The three spectra are identical when $\lambda =2m\omega /\hbar $ and this is due, as one can notice from \ref{eqt45}, to the fact that this critical value cancels the spin-orbit term.

\begin{figure}
\centering
\includegraphics[width=0.5\textwidth]{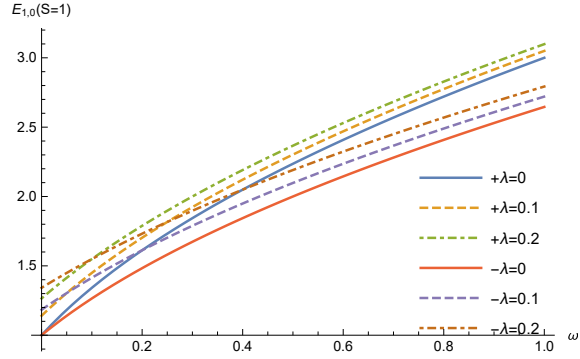}
\caption{$E_{1,0}$ versus $\omega$ for unnatural spin 1 case (the $\pm$ signs near $\lambda$ holds for $E_{\pm}$ energies) }
\label{fig4}
\end{figure}

\begin{figure}
\centering
\includegraphics[width=0.5\textwidth]{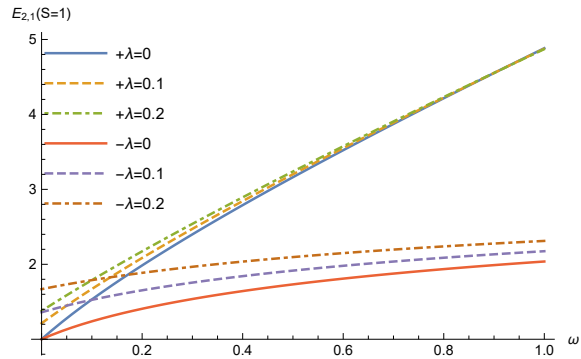}
\caption{$E_{2,1}$ versus $\omega$ for unnatural spin 1 case (the $\pm$ signs near $\lambda$ holds for $E_{\pm}$ energies) }
\label{fig5}
\end{figure}

By studying the behavior of the energies $E_{\pm }$ as a function of the frequency, we note that their dependencies are almost linear for the states $J=0$ as shown in fig\ref{fig4} for $N=1$. Whereas for the states $J\neq 0$, both energies are linear only for weak frequencies, then their behaviors differ in the high frequencies regime and we see, as in  fig\ref{fig5}, that $E_{+}$ does not cease increasing towards infinity while $E_{-}$ tends towards a constant $\sqrt{\left( N+2\right) \left( N+3\right) /J\left( J+1\right)}$.

\begin{figure}
\centering
\includegraphics[width=0.5\textwidth]{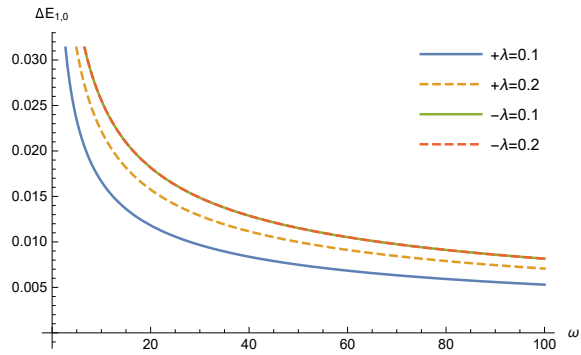}
\caption{$\Delta E_{1,0}=E(\lambda)-E(\lambda=0)$ versus $\omega$ for unnatural spin 1 case
\newline (the $\pm$ signs near $\lambda$ holds for $E_{\pm}$ energies) }
\label{fig6}
\end{figure}

\begin{figure}
\centering
\includegraphics[width=0.5\textwidth]{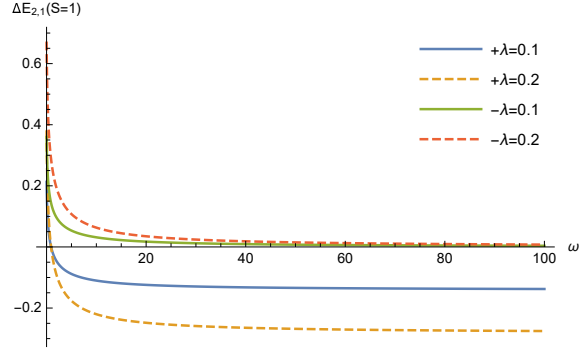}
\caption{$\Delta E_{2,1}=E(\lambda)-E(\lambda=0)$ versus $\omega$ for unnatural spin 1 case
\newline (the $\pm$ signs near $\lambda$ holds for $E_{\pm}$ energies) }
\label{fig7}
\end{figure}

If we focus on the contributions coming only from the deformation in these energies, we notice that, when $J=0$, they decrease with $\omega$ until they cancel out for both $E_{+}$ and $E_{-}$ cases (fig \ref{fig6}). On another hand, when $J\neq 0$, they decrease for the $E_{+}$ energies and tend towards the negative value $-\sqrt{J\left( J+1\right)}\lambda$, while they also decrease for the $E_{-}$ energies until becoming zero for very high frequencies (fig\ref{fig7} as an example for the level $(N,J)=(2,1)$).

We associate the energies $E_{\pm }$ with the states $L=J\pm 1$ \cite{Nedjadi941, Nedjadi942, Falek10} and we mention that the same matches appear in 2D DKP systems \cite{Falek19} but they are absent for 1D ones \cite{Boumali08, Falek17} since there is neither orbital moment nor spin in this case. We also associate the natural states with the vanishing projection of the $S=1$ case, or $J=L+0$, and this explains the fact that its spectrum \ref{eqt52} is similar to that of the $S=0$ case in \ref{eqt35} (or \ref{eqt53} and \ref{eqt54} as mentioned before) and it also explains the absence of spin--orbit coupling for this parity. By making the analogy with the case of Dirac's equation, this gives a new meaning to natural and non-natural states since the parity is defined by $\left( -1\right) ^{L}$ in relativistic quantum mechanics \cite{Bjorken64} and so natural states correspond to $L=J$ while unnatural ones are related to $L=J+1$ and to $L=(J+2)-1$.

\section{Solutions for deSitter Spaces}
\label{sec:dS}
As we have already mentioned in section \ref{sec:RV}, one only has to change the sign of the deformation parameter $\lambda$ to obtain the expressions in the dS case from those corresponding to the AdS one.

For the modified Heisenberg uncertainty relations, we transform eq.\ref{eqt4} to the following:
\begin{equation}
\Delta X_{i}\Delta P_{i}\geq \frac{\hbar }{2}\left( 1-\lambda \left( \Delta X_{i}\right) ^{2}\right)  \tag{68}  \label{68}
\end{equation}
it means that we have no minimum value for $\Delta P$ in this case as shown in fig.\ref{fig1}.

For spin 0 particle we use eq.\ref{eqt35} to write the energy eigenvalues:
\begin{equation}
E_{N,J}^{2}=m^{2}c^{4}+2\hbar m\omega c^{2}N-\lambda \hbar ^{2}c^{2}\left[
\left( N+1\right) ^{2}-J\left( J+1\right) -1\right]  \tag{69}  \label{69}
\end{equation}

\section{Conclusion}
\label{sec:CC}
In this work, we have exposed an explicit calculation of the relativistic Duffin--Kemmer--Petiau oscillator in three-dimension spaces with the presence of minimal uncertainty in momentum for anti-de Sitter model. We have used the Nikiforov--Uvarov method to solve both spin 1 and spin 0 cases and we have written the wave functions in the representation of vector spherical harmonics. For both scalar and vector bosons, we obtained the complete expressions of the eigenfunctions analytically in terms of the Jacobi polynomials. We also deduced the corresponding energy eigenvalues and we found that the deformation added a hard confinement term ($\propto N^{2}$) and a rotational term ($\propto J(J+1)$ for both spin 0 particle and natural parity states of the spin 1 case. However for unnatural parity states of the spin 1 particle, there was extra additions of a new spin-orbit contribution and a supplementary rotational term to the ones that already existed for this parity in the absence of deformation.

Moreover, in order to see the effect of the deformation on physical systems, we compared them with the experimental results from the cyclotron motion of an electron in a Penning trap and we have determined the upper bound of the minimal momentum uncertainty for AdS space. In addition, our results were tested by deducing the limit formulas of the spectra for $\lambda \rightarrow 0$ and we have obtained the results of the ordinary relativistic quantum harmonic oscillator for both scalar and vector particles.

Our results show that the spectrum of the natural states of spin 1 particle is similar to the one of the scalar particle. On the other hand, that of the non-natural states of the vector particle differs from the previous two by the presence of the spin--orbit coupling. We have associated the natural states with $L=J+0$ of the $S=1$ case and this explains the fact that its spectrum is similar to that of the case $S=0$ and also explains why there is no spin--orbit coupling for them since the projection of the spin is null in these two cases. On another side, the unnatural states are associated with both $L=J+1$ and $L=(J+2)-1$ for the $S=1$ case or when the projection of the spin in either $+1$ or $-1$; this explains the presence of the spin--orbit coupling even without deformation. This new interpretation allows us to redefine the parity with $\left( -1\right) ^{L}$ as it is normally done in relativistic quantum theory.

\section*{Acknowledgment}

This work was done with funding from the DGRSDT of the Ministry of Higher Education and Scientific Research in Algeria as part of the PRFU B00L02UN070120190003.

\end{document}